\documentstyle[11pt,epsfig]{article}
\begin{document}
\title{Expected Properties of Massive Neutrinos\\
for Mass Matrices with a Dominant Block\\
and Random Coefficients Order Unity}
\author{Francesco Vissani\footnote{e-mail: {\tt vissani@lngs.infn.it}}\\[1ex]
INFN, Laboratori Nazionali del Gran Sasso,\\
Theory Group. Assergi (AQ), Italy}
\date{}
\maketitle
\begin{abstract}
We study the class of neutrino mass matrices with 
a dominant block but unspecified ${\cal O}(1)$ 
coefficients, and scan the possible models by the help of random 
number generators. 
We discuss which are the 
most common expectations in dependence of the adjustable
parameter of the mass matrices, $\varepsilon,$ and 
emphasise an interesting sub-class of models 
that have large mixing angles for atmospheric 
and solar neutrinos, and an angle $\theta_{13}$ 
close to the experimental limit.
For those models where the lepton mass matrices 
are subject to Froggatt-Nielsen U(1) selection rules,
we show that the neutrino mixing matrix receives 
important contributions from the rotations operating 
on charged lepton sector, which increase the predicted value of the
angle $\theta_{13}$ and the ee-entry of the neutrino mass.
\end{abstract}
\parskip.4cm
A specific, simple form of neutrino mass matrix has stimulated
many studies in the last few years (see {\em e.g.}\ 
\cite{ram,sy1,altf,vi,ki,ha,sy2,ta,muha,mk}):
\begin{equation}
{\bf M}_\nu\propto
{\mathrm diag}[\varepsilon,1,1]\; {\cal O}(1)\; {\mathrm diag}[\varepsilon,1,1]
\label{e_matr}\label{1}
\end{equation}
where it is understood that $\varepsilon$ is a 
small parameter\footnote{The case $\varepsilon=1$ 
has been given theoretical support in \cite{zu}, and was 
 discussed in detail in \cite{ha} and \cite{sy2}; however, 
we regard it as an extreme case.}, 
and ${\cal O}(1)$ is the matrix of the ``coefficients 
of order unity'' (an argument 
for eq.\ (\ref{1}) with a minimum 
of theoretical assumptions is presented in \cite{mg9}).
The attempt to understand the 
gross features of 
the fermion (neutrino) mass matrices by the help of  
factor $\varepsilon$ but without stressing 
too much the role of the coefficients
is in the line of thought drawn by Froggatt and Nielsen \cite{fn}.
In their language, the matrix in eq.\ 
(\ref{1}) is characterised by the 
presence of two degenerate U(1) charges, 
and this implies that a block of the mass matrix has relatively
large entries (``dominant block''). 
Recently, the early 
suggestion \cite{fn} that the coefficients should be 
thought as random numbers was taken quite literally in 
the studies \cite{ha,sy2,muha,mk}.
We find this type of approach of interest, 
since it amounts to a  scan of the 
possible theoretical models;
the most frequent cases (in dependence on the value of $\varepsilon$)
are emphasised in this approach.
We show in the appendix our assumptions on the 
coefficients, seen as random numbers, 
and discuss related problems.

\begin{figure}[tb]
\centerline{\hskip2.5cm\epsfig{file=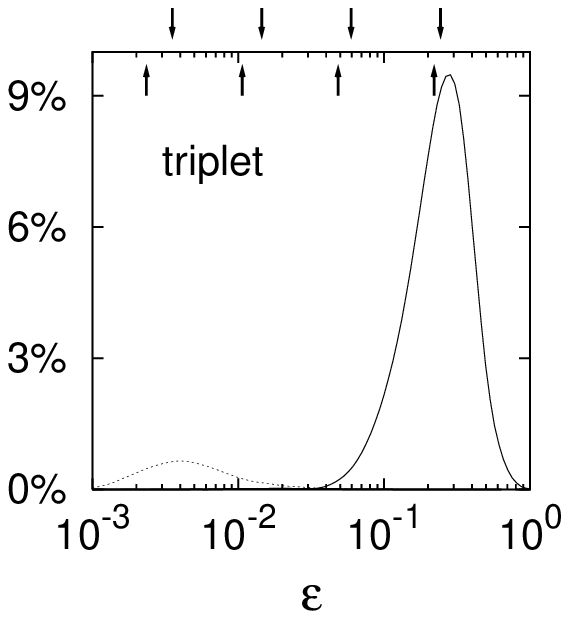}\hskip-2.8cm\epsfig{file=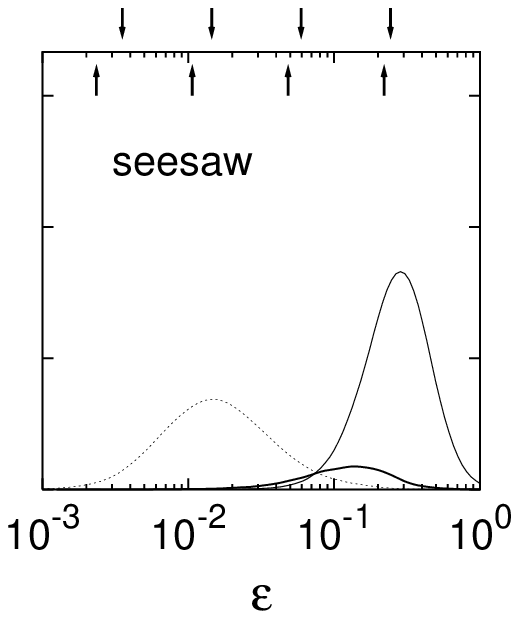}}
\caption{Percentage of phenomenologically
successful neutrino mass matrices of the type eq.\ (\ref{1})
as a function of $\varepsilon.$ 
Dashed line denotes the {\sc sma} region,
continuous thin line the {\sc lma} region, 
thick line the {\sc low} region
(last one is practically invisible for triplet case).
For orientation, we  
emphasise with arrows pointing downward the special cases
$\protect{\varepsilon}=(m_\mu/m_\tau)^{0.5,1,1.5,2}$
and with those pointing upward the cases
$\protect{\varepsilon}=(\sin\vartheta_C)^{1,2,3,4}$
($\vartheta_C=$Cabibbo angle).} 
\label{fig:perc}
\end{figure}

We begin by discussing the value of $\varepsilon$ 
putting emphasis on available neutrino oscillation data.
Postponing the interpretation of LSND indications, 
we know with good confidence that\footnote{As usual, 
${\bf M}_\nu=U^* {\mathrm diag}[m_j\times \exp(i\xi_j)] U^\dagger,$
$m_j\le m_{j+1}$ and $\Delta m^2_{ji}=m_j^2-m_i^2.$
In terms of the neutrino mixing matrix $U$ 
(such that $\nu_\ell=U_{\ell i}\nu_i,$ 
with $\ell={\rm e},\mu,\tau$ and $i=1,2,3$) 
the mixing angles $\theta_{ij},$ with values 
$0^\circ\le  \theta_{ij}\le 90^\circ,$ 
are simply $\sin\theta_{13}=|U_{\rm e3}|,$
$\tan\theta_{12}=|U_{\rm e2}/U_{\rm e1}|$ and
$\tan\theta_{23}=|U_{\rm \mu 3}/U_{\rm \tau 3}|.$} 
the atmospheric mixing angle $\theta_{23}$ 
is in the range $45^\circ\pm 10^\circ,$ while  $\theta_{13}$ belongs 
roughly to $(0-10)^\circ;$ the corresponding $\Delta m^2_{31}$ is 
in the range $(1.5-5)\times 10^{-3}$ eV$^2.$
(It is rather evident that the two informations 
on the mixing angles are in reasonable agreement with 
the mass matrix in eq.\ (\ref{1}), if $\varepsilon$
is sufficiently small). The situation with solar neutrino
observations is 
less clear. We assume that the observations 
indicate again oscillations of massive neutrinos, with
$\Delta m^2_{21}\ll \Delta m^2_{31},$
and select three regions of parameter space for further 
discussion: (1) a {\sc sma} region, namely the rectangle 
with vertices $(4\times 10^{-6},2\times 10^{-4})$ 
and $(1\times 10^{-5},3\times 10^{-3})$ in the 
$(\Delta m^2_{21}/\mbox{eV}^2,\tan^2\theta_{12})$ plane,
(2) a {\sc lma} region, namely the rectangle 
with vertices $(8\times 10^{-6},0.15)$ 
and $(3\times 10^{-4},0.75)$ in the same plane,
(3) a {\sc low} (up to quasi-{\sc vo}) region, that is
the rectangle with vertices $(6\times 10^{-10},0.3)$ 
and $(3\times 10^{-7},3).$
Indeed, since these models 
do not predict the overall mass scale, we use $\Delta m^2_{21}$
to calculate values of the {\em hierarchy factor:}
$$
h=\frac{\Delta m^2_{21}}{\Delta m^2_{31}}
$$ 
assuming that $\Delta m^2_{31}=3 \times 10^{-3}$ eV$^2,$ and 
then we compare these values of $h$ with the calculated ones. 
The percentage of mass matrices that passes the
cuts on $\theta_{23},\theta_{13},\theta_{12}$ and $h$
is shown in fig.\ \ref{fig:perc}. 
As pointed out in \cite{ha,sy2}, it is of some 
importance to distinguish
the case denoted as ``triplet'' in the figure,
when one generates a matrix with random coefficients 
in eq.\ (\ref{1}) (namely ${\cal O}(1)=R_1;$
see \cite{go} for theoretical support) 
from the case when ${\cal O}(1)=R_2 R^{-1}_3 R^t_2,$ 
which corresponds to assume that the light neutrino masses
are due to the ``seesaw'' mechanism \cite{ss}. In last case,
the random coefficients are those of the Dirac neutrino couplings, 
and of the heavy (right handed) neutrinos--$R_2$ and $R_3$ 
respectively.\footnote{In general, both contributions are 
present: ${\cal O}(1)=\sin^2\eta\, R_1 +\cos^2\eta\, R_2 R^{-1}_3 R^t_2.$
For simplicity, we will discuss only the extreme cases--triplet 
and seesaw--but it is clear that we can bias this or that value of 
$\eta$ only adding theoretical information.}
It is seen that a reasonably large number of mass matrices of the type 
eq.\ (\ref{1}) survive the cuts for certain values of $\varepsilon.$
It is rather obvious that for very large values of
$\varepsilon$ the number of successful models decreases 
due to the cut on $\theta_{13}.$ 
The lower peak height of the {\sc sma}- (even more, {\sc low}-)
curves is due to the difficulty to reduce sufficiently $h$ (which is 
alleviated in the ``seesaw'' case \cite{ha,sy2}). 
In both cases, there is a gap with relatively  
small overlap between the {\sc sma} and {\sc lma} curves.
This can be explained in the following manner:
consider the plane $(h,\theta_{12});$ as typical values,
we have $h\sim 1$ and $\theta_{12}\sim \varepsilon,$ 
but there is an interesting tail
(=a less populated region) which shrinks 
with $\varepsilon$ where  $\theta_{12}$ 
increases and $h$ decreases (the correlation being
tighter in the ``triplet'' case). For diminishing
values of $\varepsilon,$ the tail first meets 
the {\sc lma} region and only after 
moves toward the {\sc sma} region: this creates the 
gap observed in fig.\ \ref{fig:perc}.\footnote{This is the 
same argument put forward  in sect.\ 2 of ref.\ \cite{vi}.
But note that eq.\ (2.1) therein is improved 
assuming that the 33-entry is 2, not 1;
at the same time, $m_\nu\sim (\Delta m^2_{atm})^{1/2}$
has to be replaced by $m_\nu\sim (\Delta m^2_{atm})^{1/2}/2.$ 
This affects the numerical factor in front to
eqs.\ (2.2-4), and  the region in figure 1 
shifts downward by a factor of 4.\label{foo}} The different
position and height of the ``{\sc sma}-peak'' from seesaw 
to triplet case indicate just that the distributions (and tails) 
are different. 

\begin{table}[tb]
\begin{center}
\begin{tabular}{|ccc|ccc|c|}
\hline
  $Q_{\rm e}$ &  $Q_\mu$ &  $Q_\tau$ &
  $Q_{\rm e^c}$ &  $Q_{\mu^c}$ &  $Q_{\tau^c}$ &
  $\varepsilon$ (degrees) \\
\hline
3 & 0 & 0 & 3 & 2 & 0 &  $.83^\circ$ \\
2 & 0 & 0 & 4 & 2 & 0 &  $3.4^\circ$ \\
1 & 0 & 0 & 5 & 2 & 0 &  $14.^\circ$ \\
\hline
\end{tabular}
\end{center}
\caption{Three sets of 
Froggatt-Nielsen leptonic charges which
give an ascending series of $\varepsilon=(v/M)^{Q_{\rm e}}$ values 
($v/M$ is discussed in the text). We normalise to 0 
the lowest charges since we focus
only on charged lepton mass ratios; $Q_{\mu^c}=2$ is 
needed to reproduce $m_\mu/m_\tau;$ 
and $Q_{\rm e}+Q_{\rm e^c}=6$ (or nearby 
values) to reproduce  
$m_\mu/m_\tau=(m_{\rm e}/m_\tau)^{1/3}$ 
(correct at the 10 \% level). 
\label{tab:ch}}
\end{table}
Some models of this type 
(=values of $\varepsilon$) have been discussed already
in the literature: $\varepsilon=\sin^3\vartheta_C$ in \cite{ram},
$=m_\mu/m_\tau$ in \cite{sy1,sy2}, $=\sin^n\vartheta_C$ in \cite{vi},
1 in \cite{ha}. Except than in the last case, the stress was put
on the correlations of the properties of neutrinos with those
of charged fermions. However, no special emphasis has 
been put on the case:
\begin{equation}
\varepsilon=\sin\vartheta_C\approx 0.22\ \ \ \ \
\mbox{or}\ \ \ \ \ =(m_\mu/m_\tau)^{1/2}\approx 0.24,
\label{2}
\end{equation}
that we see to be quite interesting in connection with 
the {\sc lma} region (in both the ``seesaw'' 
and the ``triplet'' case). 
The case $\varepsilon=m_\mu/m_\tau \approx 0.06$
was emphasised in reference \cite{sy2}; it was shown that 
the use of U(1) selection rules {\em a la} Froggatt and Nielsen,
with charges $Q_{\rm e}=1,$ $Q_\mu= Q_\tau=0$ 
for the left fields is not in  contradiction with 
the gross features of the charged lepton spectrum, 
if at the same time one assumes
$Q_{{\rm e}^c}=2,$ $Q_{\mu^c}=1$ and $Q_{\tau^c}=0$ 
as charges for the right fields. The same can be however obtained 
with other choices of charge, as those in table \ref{tab:ch}
with a different value of the parameter that regulates all mixings 
and hierarchies, $v/M=(m_\mu/m_\tau)^{0.5}$
(note that the second model is just the one of 
Sato and Yanagida \cite{sy2}, since $Q\to 2 Q$ but also
$v/M\to  (v/M)^{1/2}$ in our table). 

We verified numerically the viability  
of the alternative choices, and found that  
increasing $Q_{{\rm e}^c}$ by 1 unity also 
leads to similar results, for slightly larger
values of $\varepsilon;$ similar conclusions even
decreasing $Q_{{\rm e}^c}$ by 1 unity 
(but worsening a bit the agreement with charged fermion
mass ratios in the first two cases of table \ref{tab:ch}): 
Nothing against large values of $\varepsilon$ 
as in eq.\ (\ref{2}) from this side. 
Indeed, it was argued already in \cite{fn} that the
value of $v/M$ should be large, 
simply in connection with the size of the
Cabibbo angle; $\sin\vartheta_C$ 
should presumably come out as the result of a fluctuation, 
if $v/M$ were much smaller.

Now we tackle one last problem, which is 
strictly linked to the 
theoretical (Froggatt-Nielsen) context. We 
start with some general consideration:
One can attempt to classify 
the models for neutrino masses, 
depending on whether the neutrino mixing matrix
\begin{equation}
U=U_E^\dagger\ U_\nu
\label{3}
\end{equation}
is {\em 1.}\ mostly due to the rotation of the 
neutral leptons $U_\nu,$ or {\em 2.}\ to 
the rotation of charged leptons $U_E,$ 
or, finally, {\em 3.}\ if both have a comparable 
importance (at least for some mixing angle). 
It is easy to argue for the first possibility 
(at least in words); neutrinos look special, 
charged leptons could always entail small mixing, 
and thence $U_\nu$ could play an overwhelming role in $U.$
As an example of the second possibility, we quote the ``lopsided'' 
models described in \cite{ba}.
Finally, as an important instance of the last possibility, 
$U_E$ and $U_\nu$ are typically of 
comparable importance for $U$ in 
the U(1) approach of Froggatt and Nielsen \cite{fn}, 
since they are controlled by the \underline{same} charges,
namely, those of the left leptonic fields (note, however, that 
a partial degeneracy in the neutrino mass matrix 
might enhance the role of $U_\nu$). 
Strictly speaking, most of the considerations above apply 
to the first case, in that (following previous studies)
we simply ignored the role of the mixing due to 
the charged leptons ($U_E\approx 1\!\!1$). 
However, since the approach with random numbers 
is naturally (though not unavoidably) connected with the use of
U(1) selection rules for fermion mass matrices, and since the 
existing studies of neutrino mass matrices of this 
type \cite{sy2,ha,muha,mk} do not take into account the point,
we decided to investigate what is the effect of $U_E$ 
in models with U(1) selection rules.
Thence, we additionally generated
the random mass matrices $R_0$ 
as coefficients of the charged lepton mass matrix 
${\bf M}_E=U_E^*\ {\mathrm diag}[m_\ell]\ V_E:$
$$
{\bf M}_E
\propto {\mathrm diag}\left[\left(\frac{v}{M}\right)^{Q_{\ell'}}\right]\; 
{\cal O}(1)\; 
{\mathrm diag}\left[\left(\frac{v}{M}\right)^{Q_{\ell^c}}\right]
$$ 
(where $\ell,\ell'={\rm e},\mu,\tau$)
enforcing the values of the U(1) 
charges in tab.\ \ref{tab:ch}
to specify the models fully. 
Only if the mass ratios $(m_{\rm e}/m_\tau)$ and 
$(m_{\mu}/m_\tau)$ are reproduced within 30 \%,
we calculate $U_E$ and estimate its effect on $U$
in eq.\ (\ref{3})
(it is important to implement such a condition 
to gauge out cases when the mixing angles 
come out artificially large). 

\begin{table}[tb]
\begin{tabular}{|l|ccccc|rrr|}
\hline
$.83^\circ$       & $45-\theta_{23}\!\!\!\!$ 
& $\theta_{13}$ & $\theta_{12}$ & $h$ & 
$m_{\rm ee}/10^{-4}$ & $\!\!${\sc sma}$\!$ 
& $\!\!${\sc lma}$\!$ & $\!\!${\sc low}$\!$\\
\hline
{\tt$\!$t,w/$\!$o}$\!$ & $\!\pm\! 12$ 
      & $\!.37 \!\pm\! .19$ 
      & $\!1.0 \!\pm\! 1.4$ 
      & $\!.35 \!\pm\! .26$           
      & $\!1.4 \!\pm\! 3.3$ 
      &.15
      &.00
      &.01	
\\
{\tt$\!$t,w}   & $ \!\pm\! 23$ 
      & $\!.70 \!\pm\! .33$ 
      & $\!1.2 \!\pm\! 1.4$ 
      & $\!.35 \!\pm\! .26$           
      & $\!2.9 \!\pm\! 1.7$ 
      &.04 
      &.00
      &.00
\\     
{\tt$\!$s,w/$\!$o}$\!$ & $ \!\pm\! 17$ 
      & $\!.52 \!\pm\! .29$ 
      & $\!1.3 \!\pm\! 1.7$ 
      & $\!.12 \!\pm\! .16$           
      & $\!1.4 \!\pm\! 1.3$ 
      &2.0
      &.00
      &.02
\\     
{\tt$\!$s,w}   & $ \!\pm\! 21$ 
      & $\!.79 \!\pm\! .41$ 
      & $\!1.5 \!\pm\! 1.7$ 
      & $\!.12 \!\pm\! .16$           
      & $\!2.9 \!\pm\! 2.4$ 
      &.92
      &.00
      &.01
\\
\hline
\hline
$3.4^\circ$       &  &  &  &  & 
$m_{\rm ee}/10^{-3}$ &  & & \\
\hline
{\tt$\!$t,w/$\!$o}$\!$ & $\!\pm\! 12$ 
      & $\!1.5 \!\pm\! 0.8$ 
      & $\!3.8 \!\pm\! 3.8$ 
      & $\!.35 \!\pm\! .26$           
      & $\!2.4 \!\pm\! 0.6$ 
      &.01
      &.48
      &.01
\\
{\tt$\!$t,w}   & $\!\pm\! 23$ 
      & $\!2.9 \!\pm\! 1.4$ 
      & $\!4.6 \!\pm\! 3.8$ 
      & $\!.35 \!\pm\! .26$           
      & $\!4.9 \!\pm\! 2.9$ 
      &.00 
      &.12
      &.00
\\     
{\tt$\!$s,w/$\!$o}$\!$ & $\!\pm\! 17$ 
      & $\!2.1 \!\pm\! 1.2$ 
      & $\!5.0 \!\pm\! 5.0$ 
      & $\!.12 \!\pm\! .16$           
      & $\!2.3 \!\pm\! 2.1$ 
      &.58
      &.18
      &.27
\\     
{\tt$\!$s,w}   & $\!\pm\! 21$ 
      & $\!3.3 \!\pm\! 1.7$ 
      & $\!5.7 \!\pm\! 5.1$ 
      & $\!.12 \!\pm\! .16$           
      & $\!4.9 \!\pm\! 4.0$ 
      &.24
      &.15
      &.07
\\
\hline
\hline
$14.^\circ$    
& & & & & 
$m_{\rm ee}/10^{-2}$ &  & & \\
\hline
{\tt$\!$t,w/$\!$o}$\!$ & $\!\pm\! 12$ 
      & $\!6.2 \!\pm\! 3.2$ 
      & $\!12.5 \!\pm\! 8.4$ 
      & $\!.36 \!\pm\! .26$           
      & $\!4.0 \!\pm\! 0.9$ 
      &.00
      &9.1
      &.00
\\
{\tt$\!$t,w}   & $ \!\pm\! 23$ 
      & $\!11.8 \!\pm\! 5.6$ 
      & $\!16.3 \!\pm\! 9.3$ 
      & $\!.36 \!\pm\! .26$           
      & $\!7.9 \!\pm\! 4.6$ 
      &.00 
      &9.9
      &.00
\\     
{\tt$\!$s,w/$\!$o}$\!$ & $ \!\pm\! 17$ 
      & $\!8.7 \!\pm\! 4.6$ 
      & $\!17.1 \!\pm\! 12.3$ 
      & $\!.13 \!\pm\! .17$           
      & $\!3.7 \!\pm\! 3.1$ 
      &.03
      &4.8
      &.32
\\     
{\tt$\!$s,w}   & $ \!\pm\! 21$ 
      & $\!13.1 \!\pm\! 6.6$ 
      & $\!20.0 \!\pm\! 12.6$ 
      & $\!.13 \!\pm\! .17$           
      & $\!7.6 \!\pm\! 5.9$ 
      &.01
      &2.0
      &.04
\\
\hline
\end{tabular}
\caption{Calculated neutrino properties for 
the U(1) models of tab.~\ref{tab:ch},  
in the cases with triplet or 
seesaw ({\tt t} and {\tt s} resp.) mass mechanism, 
and with or without the account of 
the lepton mixing matrix $U_E$ ({\tt w} and {\tt w/$\!$o} resp.).
The 3 parts of the table correspond to the models 
defined in tab.\ \ref{tab:ch} (in the left-upper corners, the
values of $\varepsilon$ in degrees).
All angles in the table are in degrees, while
{\sc sma, lma} and {\sc low} denote percentages
(cuts as in fig.\ \ref{fig:perc}).\label{tab:cfr}}
\end{table}

The results are presented in table \ref{tab:cfr}. 
The spread\footnote{The spread $\delta x$ on the quantity $x$
is calculated as $\delta x^2=\sum_i^N \left(x_i-\langle x\rangle\right)^2/N,$
where $x_i$ is the result of the $i^{th}$ simulation,
$\langle x\rangle=\sum_i^N x_i/N$ is the average,
and $N=10^6.$} 
in $\theta_{23}$ does not change much 
with $\varepsilon,$ however, it increases significantly 
with the inclusion of $U_E$ effects; it is a pity that 
these simple models are unable to give an indication on
the size of the deviation of $\theta_{23}$ from $45^\circ,$
one of the most interesting quantities to 
be searched in future experiments.
The average values of $\theta_{12}$ and $\theta_{13}$ 
also increase with the inclusion of $U_E.$ 
If $U_E$ effects are not included, $\theta_{13}$ is on average smaller
than $\varepsilon$ \cite{sy2}; the reason 
is just the numerical factor $\sim 2$ discussed in footnote \ref{foo}.
The angle $\theta_{13}$ is rather close to 
$\varepsilon$ when these effects are included, and close to
the experimental limit when $\varepsilon$ is large. 
Note that even a seemingly modest
increase in $\theta_{13},$ say by a factor of 2
(table \ref{tab:cfr}, triplet case) 
is an important message for the searches 
of $\nu_{\rm e}$ appearance in terrestrial experiments, 
since the probabilities of conversion
in vacuum depend strongly on this parameter:
$P(\nu_\mu\to\nu_{\rm e})\propto(\theta_{13})^2.$ 
In tab.\ \ref{tab:cfr} we also present the 
value of the following quantity:
$$
m_{\rm ee}=\frac{|({\bf M}_\nu)_{\rm ee}|}{(\Delta m^2_{31})^{1/2}}
$$
These calculations show that, for the larger value
of $\varepsilon$ considered, the ee-element of the mass matrix
$|({\bf M}_\nu)_{\rm ee}|$ can reach the several meV level. 
This is tantalisingly close to the expected 
sensitivity of the next generation neutrinoless 
double beta decay experiments, namely $10-20$ meV.

Let us summarise and discuss the results.
Neutrino masses are expected to be non-zero, and indeed
the atmospheric neutrino mass scale $\sim 50-60$ meV is very
well compatible with the ideas of grand unification, but the 
size of the mixing angles is a puzzle, in particular 
the strong indication that at least one of them is 
large--maybe maximal. 
We discussed a class of neutrino mass models (eq.\ (\ref{1}))
that are inspired to the principle that all 
elements of the mass matrices
(seen as fundamental quantities) should be equally large, 
unless explicitly suppressed.
We emphasised the type of models that are phenomenologically 
successful, and discussed the stability in the choice of coefficients
by the help of random number generators (fig.\ \ref{fig:perc}). 
We discussed also the connection
with known properties of the charged fermions (and, in particular,
leptons), and remarked that large values of the ``order parameter''
$\varepsilon$ (see eq.\ (\ref{2})) are quite interesting in connection 
with the {\sc lma} solution of the solar neutrino problem. 
In present context, the triplet mechanism 
is not disfavored in comparison with
the seesaw mechanism for neutrino mass generation;
one could argue instead that the triplet mechanism is more predictive,
for the simple reason that it favors 
solar neutrino solutions with as large $h$ as possible.
In close observance to the  Froggatt-Nielsen approach, we also
pointed out the relevance of the mixing due to charged 
leptons for the neutrino mixing; this effect 
unfortunately renders almost random the expected
value of the parameter $\theta_{23}$ in these models, 
though the maximal mixing $\theta_{23}=45^\circ$ is 
the most likely possibility.
Large values of the mixing angle 
$\theta_{13},$ and perhaps an observable 
$|({\bf M}_\nu)_{\rm ee}|$ can be naturally incorporated in the model,  
if the parameter $\varepsilon$ is large as in eq.\ (\ref{2}).

\vskip.5cm
\noindent{I thank W.\ Buchm\"uller, D.\ Delepine, A.\ Grillo,
P.\ Lipari, H.\ Nunokawa, G.\ Senjanovi\'c, M.\ Tanimoto and 
especially Z.G.\ Berezhiani for clarifying discussions.
After these calculations were completed, 
we learned of an interesting  study \cite{st} of
illustrative cases of the mixing matrices 
$U_E$ and $U_\nu;$ when comparison is
pertinent, our conclusions agree with theirs.}
\vskip.5cm
\noindent\underline{\it Appendix: What is a Random Coefficient?}
\vskip.3cm

\noindent In the spirit of the approach, the 
random coefficients of the mass matrices should 
not be enumerated among the parameters;
however, their form has to be fixed
in order to perform the scan of possible models.
To give an idea of the effect of changing the generator,
we compare the percentage of successful models in 
({\sc sma},{\sc lma},{\sc low})-regions for the model of 
Sato and Yanagida (seesaw case, $U_E$ 
included--tab.\ \ref{tab:cfr}, middle part, last line). 
Variations amounts to a factor of 
a few in most extreme cases:
$(0.24,0.16,0.11)$~\% for $Z_{0},$
$(0.42,0.22,0.11)$~\% for $Z_1.$ 
We denote with $Z_\delta$ the numbers in the complex plane
with random phase, and with modulus 
uniformly distributed in an interval 
$[-\delta,\delta]$ around 1 ($Z_1$ is a circle in 
the complex plane, $Z_0$ the circumference).
Following \cite{sy2}, we adopt in our study 
the choice $Z_{\delta},$ with $\delta=0.2,$
clearly consistent with the view that the 
coefficients are ``order unity''.

There is another point of ambiguity: 
How to treat a symmetric random mass matrix $R$
(indeed, the matrices
$R_1$ and $R_3$ described in the main text
are symmetric).
One possibility is to generate just 
the elements $R_{ij}$ with $i\ge j,$
and then set $R_{ji}=R_{ij};$ another one is to 
generate the full matrix, 
and then replace $R_{ij}$ and $R_{ji}$ by 
their average.\footnote{These 
are clearly different prescriptions; for instance, 
the sum of 2 numbers uniformly distributed is not 
uniformly distributed.}
Following the previous studies, we adopted the first 
prescription, and noted that the second prescription leads to 
results that differs little from those in tab.\ \ref{tab:cfr}
(in the same case of previous paragraph);
$\theta_{23}$ is the same; 
$\theta_{13}= (3.2\pm 1.6)^\circ$,
$\theta_{12}=(5.6\pm 5.0)^\circ,$
$h=.11\pm.15,$
$m_{\rm ee}=(4.7 \pm 3.9)\times 10^{-3};$
finally, the {\sc sma},{\sc lma} and {\sc low} 
success percentages are $0.26,0.10$ and $0.08$~\%. 

Now, a delicate aspect;
in this work (following previous studies) we assumed  that all 
coefficients ${\cal O}(1)$ are distributed in the 
same manner, but {\em a priori} it is unclear whether 
this assumption is fair. 
Indeed, if terms higher order in $v/M$
arise through the exchange of virtual particles as 
suggested in \cite{fn}, there might be not only a piling up of 
$(v/M)$'s
but also of the coefficients themselves.\footnote{Even if the
``fundamental'' ${\cal O}(1)$ 
coefficients had exactly modulus 1 but different phases,
the exchange of different virtual states leads to 
interference, and thence to a different distribution of the 
``effective'' ${\cal O}(1)$ coefficients.} 
We do not want to deny the interest
of this simplifying assumption (that we adopt in this study), 
but we point out a risk of minor reliability of the 
predictions which depend essentially on higher order terms. From 
this point of view, the prediction 
of a relatively large value of $h$ and of $\theta_{23}$
are more reliable than those on the 
other mixing angles; the one on $|({\bf M}_\nu)_{\rm ee}|$ 
has the greatest dependence on this crucial assumption, and might
be considered, then, even less reliable.

A final warning, of more general nature:
Mass matrices with random coefficients can 
help to emphasise certain possibilities that 
are compatible with present (lack of) information, but
one should be careful to interpret the results 
of this type of calculations 
in ``probabilistic'' term. 
Indeed, if, for consistence, 
the absolute scale of the neutrino 
mass matrix was also let fluctuate; or also, if one required
that the (masses of the) charged leptons were reproduced
within experimental errors;
{\em etc.}; the probability of success of 
these attempts would have been practically zero.
Similarly, if a theory of the coefficients order unity were 
given, any ``statistical'' consideration (like the present ones) 
would have been much less relevant.

\end{document}